\documentclass[conference]{IEEEtran}

\usepackage[utf8]{inputenc}
\usepackage[T1]{fontenc}

\usepackage{amsmath}
\usepackage{amssymb}
\usepackage{booktabs}
\usepackage{dsfont}
\usepackage{mdframed}
\usepackage[frozencache,cachedir=minted-cache]{minted}
\usepackage{tabularx}
\usepackage[svgnames]{xcolor}
\usepackage{tikz}
\usetikzlibrary{positioning}

\newcommand{\F}{\ensuremath{\mathds{F}}}

\makeatletter
\renewenvironment{minted@colorbg}[1]{
\setlength{\fboxsep}{\z@}
\def\minted@bgcol{#1}
\noindent
\begin{lrbox}{\minted@bgbox}
\begin{minipage}{\linewidth}}
{\end{minipage}
\end{lrbox}%
\colorbox{\minted@bgcol}{\usebox{\minted@bgbox}}}
\makeatother

\newminted[cppcode]{cpp}{fontsize=\scriptsize,frame=leftline,framesep=3pt,framerule=2pt,rulecolor=\color{LightGray},bgcolor=LightGray!30!white,linenos=true,numbersep=1pt}
\newminted[pycode]{python}{fontsize=\scriptsize,frame=leftline,framesep=3pt,framerule=2pt,rulecolor=\color{LightGray},bgcolor=LightGray!30!white,linenos=true,stepnumber=1,numbersep=1pt}

\setmintedinline{fontsize=\small}

\mdfsetup{leftmargin=0pt,rightmargin=0pt,innertopmargin=0pt,innerbottommargin=0pt,innerleftmargin=0pt,innerrightmargin=0pt,leftline=false,rightline=false,topline=false,bottomline=false,skipabove=3pt,skipbelow=2pt}

\newcommand{\ilcmd}[1]{{\small`\texttt{#1}'}}

\title{The EPFL Logic Synthesis Libraries}
\author{%
  \IEEEauthorblockN{Mathias Soeken, Heinz Riener, Winston Haaswijk,
     Eleonora Testa, \\ Bruno Schmitt, Giulia Meuli, Fereshte Mozafari, Siang-Yun Lee, \\Alessandro Tempia Calvino, Dewmini Sudara Marakkalage, Giovanni De Micheli}
  \IEEEauthorblockA{%
    Integrated Systems Laboratory, EPFL, Lausanne, Switzerland \\[4pt]
    \textit{https://github.com/lsils/lstools-showcase}$^\star$
  }
  \thanks{\hrulefill\par $^\star$This version of the paper discusses \emph{alice} v0.3, \emph{mockturtle} v0.2, \emph{kitty} v0.7, \emph{lorina} v0.2, \emph{percy} v0.1.2, \emph{bill} v0.1, and \emph{easy}, \emph{caterpillar}, and \emph{angel} without version numbers.\medskip}
}

\IEEEoverridecommandlockouts
\begin{document}

\maketitle

\begin{abstract}
  We present a collection of modular open source C++ libraries for the
  development of logic synthesis applications.  These libraries can be
  used to develop applications for the design of classical and
  emerging technologies, as well as for the implementation of quantum
  compilers.  All libraries are well documented and well tested.
  Furthermore, being header-only, the libraries can be readily used as
  core components in complex logic synthesis systems.
\end{abstract}

\section{Introduction}
Many problems in logic synthesis are solved by combining a set of
common techniques in an efficient way.  In this paper, we present a
collection of modular open source C++-14 and C++-17 libraries that
provide efficient implementations of common reappearing logic
synthesis tasks.  Each library targets one general aspect:
\emph{alice} eases the implementation of user interfaces and their
integration in scripting languages; \emph{mockturtle} provides generic
synthesis and optimization algorithms for logic networks;
\emph{lorina} parses logic and networks in various representation
formats; \emph{kitty} provides an effective way for explicit
representation and manipulation of Boolean functions; \emph{bill}
serves as an integration layer for symbolic reasoning engines;
\emph{percy} synthesizes optimum logic networks; and \emph{easy}
represents and synthesizes exclusive-or sum-of-product forms.
In addition, the collection is augmented by three libraries targeting
applications in quantum compilation and based on some of the
aforementioned libraries:
\emph{tweedledum} is a generic quantum compilation library with
synthesis, optimization, and mapping algorithms;
\emph{caterpillar} is a quantum circuit synthesis library targeting
fault-tolerant quantum computing; and \emph{angel} specializes in
quantum state preparation.

The libraries are well documented and well tested.  Being header-only
and requiring no strong dependencies such as Boost, the libraries can
be easily integrated into existing and new projects.  The developer
saves time by not having to re-implement common core components, and
can focus on tackling more complex logic synthesis problems. Prominent 
examples of external projects using these libraries include \emph{fiction},
a framework for design automation of field-coupled nanotechnologies~\cite{fiction},
\emph{LSOracle}, a framework for efficient logic manipulation using 
different logic optimizers~\cite{LSOracle}, and \emph{Qiskit}, a 
framework for quantum computing~\cite{Qiskit}.
A summary of all libraries presented in this paper is collected in a
``showcase''
repository,\footnote{https://github.com/lsils/lstools-showcase} which
contains links to all library repositories, and several examples in
which one or more libraries are used.

In the following, we first describe the main features and design
decisions of the individual libraries.\footnote{We refer the readers
to the dedicated paper for \emph{tweedledum} \cite{schmitt2022} and 
skip the description for this library in this paper.} We then present a showcase
example, \emph{exactmine}, which uses four of the libraries, \emph{kitty},
\emph{alice}, \emph{lorina} and \emph{percy}, to mine
optimum networks for truth tables.  The example integrates truth table
extraction from logic networks, NPN classification, and a user
interface that can be accessed from various programming languages in
less than 200 lines of code.\footnote{195 lines were counted in commit
\emph{1549bcc}, excluding white space and comments.}

\section{alice: A command shell library}
The C++ library \emph{alice} helps to create shell interfaces.  Users
can enter commands that interact with internal data structures.  The
interface supports standard shell features such as command and file
auto-completion as well as a command history. Combining several
commands allows users to create synthesis scripts.  One can write
\emph{alice} programs in a way that separates the core of a library
from the shell interface.  Inside the default shell interface, only
simple scripts in terms of sequences of commands without control flow
are supported.  However, \emph{alice} shell interfaces can be
automatically compiled into Python modules and C libraries.  Exposing
the shell interface to a Python module offers to use the various
modern Python libraries and frameworks for scripting, including data
processing (e.g., \emph{pandas}), plotting (e.g., \emph{matplotlib}),
and interactive notebooks (e.g., \emph{jupyter}).  Export to a C
library makes the developer's C or C++ library readily accessible from
many other programming languages such as JVM-based languages (e.g.,
Java), .NET-based languages (e.g., C\#), or Tcl, a scripting language
that is included in many commercial electronic design automation
tools.

\subsection*{Example}
In the remainder of this section, we describe \emph{alice} by means of
a running example, in which we write a mini shell interface for the
logic synthesis framework ABC~\cite{BM10}.\footnote{The complete
example can be found in the showcase repository.} ABC already comes
with a shell interface which allows sessions as the following.  We
like to point out that no modification to the source code of ABC were
necessary for the implementation of the example.

\begin{footnotesize}
\begin{verbatim}
abc> %read file.v
abc> %ps
function : PI = 4   PO = 3   FF = 0   Obj = 4
abc> %blast
abc> &ps
function : i/o = 23/12   and = 25   lev = 12
abc> &syn3
abc> &ps
function : i/o = 23/12   and = 21   lev = 10
abc> &w file.aig
\end{verbatim}
\end{footnotesize}

In that session, first a word-level design written in Verilog is
parsed, statistics about it are printed, before it is translated into
an And-inverter graph (AIG).  Statistics about the AIG are printed,
before it is optimized, and the statistics of the optimized network
are printed.  Finally, the optimized AIG is written into a file.

The example interacts with two different data structures: word-level
designs (manipulated by commands prefixed with a \ilcmd{\%}) and AIGs
(manipulated by commands prefixed with a \ilcmd{\&}).  In
\emph{alice}, different data structures are organized inside stores.
Each store can contain several instances of a data structure, and if a
store is not empty it points to some \emph{current} store element.  A
macro API in \emph{alice} makes it easy to register such stores:

\begin{mdframed}
\begin{cppcode}
ALICE_ADD_STORE(Gia_Man_t*, "aig", "a", ...)
ALICE_ADD_STORE(Wlc_Ntk_t*, "wlc", "w", ...)
\end{cppcode}
\end{mdframed}

A store is defined by means of the data type of elements it contains.
All store related functionality is associated with library code by
referring to the store type.  No modification of the library code and
no wrappers around library types are needed.  The second and third
argument are command line flags to address specific stores in the
command.  For example, in order to print statistics about the current
AIG one writes \ilcmd{ps -{}-aig}, or \ilcmd{ps -a}; and to print
statistics about the current word-level design one writes \ilcmd{ps
  -{}-wlc} or \ilcmd{ps -w}.  The command \ilcmd{ps} is one of several
generic store commands, which are by default contained in the shell
interface.  Macros are used to associate functionality with these
commands.  The following code implements the behavior of the
\ilcmd{ps} command for AIGs and word-level networks:

\begin{mdframed}
\begin{cppcode*}{firstnumber=3}
ALICE_PRINT_STORE_STATISTICS(Gia_Man_t*, os, aig)
{
  Gps_Par_t params{};
  Gia_ManPrintStats(aig, &params);
}

ALICE_PRINT_STORE_STATISTICS(Wlc_Ntk_t*, os, ntk)
{
  Wlc_NtkPrintStats(ntk, 0, 0, 0);
}
\end{cppcode*}
\end{mdframed}

File I/O works by first defining a file type, and then associating
store types to the file type.  The next code shows how to read a
Verilog file into a word-level network and how to write an AIG into an
Aiger file, a commonly used file format to store AIGs.

\begin{mdframed}
\begin{cppcode*}{firstnumber=13}
ALICE_ADD_FILE_TYPE(verilog, "Verilog")
ALICE_ADD_FILE_TYPE(aiger, "Aiger")

ALICE_READ_FILE(Wlc_Ntk_t*, verilog, filename, cmd)
{
  return Wlc_ReadVer((char*)filename.c_str(), nullptr);
}

ALICE_WRITE_FILE(Gia_Man_t*, aiger, aig, filename, cmd)
{
  Gia_AigerWrite(aig, (char*)filename.c_str(), 1, 0);
}
\end{cppcode*}
\end{mdframed}

That code adds two commands to the \emph{alice} shell:
\ilcmd{read\_verilog -w} and \ilcmd{write\_aiger -a}.  In this
example, where Verilog can only be read into word-level netlists, and
AIGs can only be written into Aiger files, one can omit the flags to
the store.

Before we discuss how to add an arbitrary command, we discuss the
generic store command \ilcmd{convert} that is used to convert one
store element into another.  In our example, we provide a conversion
from word-level networks into AIGs to implement ABC's
command \ilcmd{\%blast}.

\begin{mdframed}
\begin{cppcode*}{firstnumber=25}
ALICE_CONVERT(Wlc_Ntk_t*, ntk, Gia_Man_t*)
{
  return Wlc_NtkBitBlast(ntk, nullptr);
}
\end{cppcode*}
\end{mdframed}

That code adds a flag \ilcmd{-{}-wlc\_to\_aig} to the command
\ilcmd{convert}.  Aliases help to simplify commands, e.g., the
following two \emph{alice} commands

\begin{footnotesize}
\begin{verbatim}
alias blast "convert --wlc_to_aig"
alias "(\w+) > (\w+)" "convert --{}_to_{}"
\end{verbatim}
\end{footnotesize}

\noindent allow us to write either \ilcmd{blast} or \ilcmd{wlc > aig} as
an alternative for \ilcmd{convert -{}-wlc\_to\_aig}.

Finally, we add a custom command.  For custom commands, \emph{alice}
supports argument parsing, argument validation, and logging.  In the
running example, we show how to implement a simple command
\ilcmd{syn3} that takes no parameters.  We refer to the documentation
for more complex implementations of custom commands.

\begin{mdframed}
\begin{cppcode*}{firstnumber=29}
ALICE_COMMAND(syn3, "Optimization", "Optimizes AIG")
{
  auto& aig = store<Gia_Man_t*>().current();
  auto aig_new = Gia_ManAigSyn3(aig, 0, 0);
  abc::Gia_ManStop(aig);
  aig = aig_new;
}
\end{cppcode*}
\end{mdframed}

Note that the store is accessed using a type parameter.  Finally, the
\emph{alice} program is completed by the following statement:

\begin{mdframed}
\begin{cppcode*}{firstnumber=36}
ALICE_MAIN(abc2)
\end{cppcode*}
\end{mdframed}

Together with some include statements and a namespace
definition, these 36 lines of code are sufficient to replicate the
shell session from the beginning of the section, which in the
\emph{alice} shell looks as follows:

\begin{footnotesize}
\begin{verbatim}
abc2> read_verilog file.v
abc2> ps -w
function : PI = 4   PO = 3   FF = 0   Obj = 4
abc2> convert --wlc_to_aig
abc2> ps -a
function : i/o = 23/12   and = 25   lev = 12
abc2> syn3
abc2> ps -a
function : i/o = 23/12   and = 21   lev = 10
abc2> write_aiger file.aig
\end{verbatim}
\end{footnotesize}

\subsection*{Logging}
One feature of \emph{alice} is that each command can log data in JSON
format into a log file.  This eases the retrieval of data and avoids
cumbersome program output parsing using regular expressions and string
manipulation.  Each custom command can control which data should be
logged, but also some generic store commands can be configured to log.
The following code implements logging for \ilcmd{ps -a}.

\begin{mdframed}
\begin{cppcode}
ALICE_LOG_STORE_STATISTICS(Gia_Man_t*, aig)
{
  return {
    {"name", Gia_ManName(aig)},
\end{cppcode}
\end{mdframed}
\begin{mdframed}
\begin{cppcode*}{firstnumber=5}
    {"inputs", Gia_ManPiNum(aig)},
    {"outputs", Gia_ManPoNum(aig)},
    {"nodes", Gia_ManAndNum(aig)},
    {"levels", Gia_ManLevelNum(aig)}};
}
\end{cppcode*}
\end{mdframed}

Calling the previous session using the log option \ilcmd{-l logfile}
will create a JSON file \emph{logfile} that contains a JSON array with
7 entries, one for each executed command.  The log entry for the first
\ilcmd{ps -a} command will look as follows:

\begin{footnotesize}
\begin{verbatim}
{"command":"ps -a", "inputs":23, "levels":12,
 "name":"function", "nodes":25, "outputs":12,
 "time":"..."}
\end{verbatim}
\end{footnotesize}

\subsection*{Python interface}
The very same code that was implemented in the example to develop the
stand-alone shell interface can be compiled into a Python module, by
just changing some compile definitions.  Each command then corresponds
to a Python function, whose arguments are according to the command
arguments and whose return value is according to the log data it
produces.  The following example illustrates the analogy:

\begin{mdframed}
\begin{pycode}
import abc2

abc2.read_verilog(filename="file.v")
abc2.ps(wlc=True)
abc2.convert(wlc_to_aig=True)
gates_before = abc2.ps(aig=True)["nodes"]
abc2.syn3()
gates_after = abc2.ps(aig=True)["nodes"]

if gates_after < gates_before:
  abc2.write_aiger(filename="file.aig")
\end{pycode}
\end{mdframed}

As can be seen, Python can be used for scripting around the shell.
That code corresponds to the previous example session, but it only
writes the AIG into a file, if the optimization step \ilcmd{syn3} leads
to an improvement.  More involved examples, including examples in C\#
and Scala, are in the showcase.

\section{mockturtle: A logic network library}
The C++ library \emph{mockturtle} provides generic logic synthesis
algorithms and logic network data structures.~\cite{RienerTH+19} The main philosophy of
\emph{mockturtle} is that all algorithms are generic in the sense that
they are independent from the implementation of the logic network data
structure. In order to achieve this, \emph{mockturtle} makes use of
concept-based design using some modern C++-17 language features. The
library design is based on a principle of four layers that depend on each
other in a linear order, with the most fundamental layer being the
\emph{network interface API}, followed by \emph{algorithms}, followed
by \emph{network implementations}, and finally a layer for
\emph{performance tweaks} on the top. Note how this order of layers
reflects the algorithm’s independence of the network data structures.

Key is to not use dynamic polymorphism, as this would (i) harm
performance and (ii) add an unnecessary dependency on the
implementation of network data structures, e.g., by extending an
abstract base class. The base of the framework is provided by the
\emph{network interface API}. It defines naming conventions for types
and methods in classes that implement network
interfaces, of which most are optional. The network interface API does
not provide any implementations for a network though. Instead, it
helps to agree on names; and if developers follow these names in their
own implementation of logic network data structures, then these can be
used together with the logic synthesis algorithms developed in
\emph{mockturtle}.

For example, the network interface API suggest names like
\mintinline{cpp}{size}, \mintinline{cpp}{is_pi}, and
\mintinline{cpp}{is_constant} to implement functionality that returns
the number of nodes in a network, and check whether a node is a
primary input or a constant, respectively. The algorithm
\mintinline{cpp}{cut_enumeration}, which can enumerate priority
$k$-cuts~\cite{CWD99} in a network, requires these functions together
with \mintinline{cpp}{get_node}, \mintinline{cpp}{node_to_index},
\mintinline{cpp}{foreach_node}, and
\mintinline{cpp}{foreach_fanin}. Now this algorithm can be called with
any network implementation as long as it implements these
functions---independent of the underlying gate library and
implementation details. The following code shows how to call the
algorithm and then print all cuts for all nodes on some network
\mintinline{cpp}{ntk}.

\begin{mdframed}
\begin{cppcode}
auto cuts = cut_enumeration(ntk);

ntk.foreach_node([&](auto node) {
  std::cout << cuts.cuts(ntk.node_to_index(node)) << "\n";
});
\end{cppcode}
\end{mdframed}

The third layer consists of actual network implementations for some
network types that implement the network interface API, e.g.,
And-inverter graphs, Majority-inverter graphs, XOR-majority graphs, or
$k$-LUT networks. Static compile-time assertions in the algorithms are
guaranteeing that compilation succeeds only for those network
implementations that do provide all required types and methods. We
make use of static inheritance in so called \emph{views} to extend or
modify a network implementation’s functionality. For example, if
\mintinline{cpp}{Network} is a network type, then
\mintinline{cpp}{topo_view<Network>} is also a network type that
guarantees that nodes are visited in topological order. Other views
exist in the library and they can be composed arbitrarily. Since the
composition is based on static inheritance, it does not add any
runtime overhead.

Finally, to guarantee a fast implementation of algorithms, we use
static conditional checks on the network type to execute specific
implementations based on the network data structure’s properties. For
example, cut enumeration requires to enumerate over all elements in
all sets of the children of a node. If the network data structure is
heterogeneous and has a different number of fanins for internal nodes,
then this loop must be computed dynamically. However, if the network
data structure is homogeneous (e.g., in the case of an And-inverter
graph, each node has two children), then two nested for-loops would
simply suffice. The library \emph{mockturtle} implements such
performance tweaks in various places without losing the generic
interface.

\emph{mockturtle} implements a variety of logic synthesis and 
verification algorithms, including
\begin{itemize}
  \item \emph{Decomposition}: disjoint support decomposition,
  Shannon decomposition, bi-decomposition~\cite{Mishchenko01}, etc.
  \item \emph{Network information extraction}: network simulation,
  simulation pattern generation~\cite{lee2021}, don't-care extraction,
  cut enumeration, etc.
  \item \emph{Logic restructuring and optimization}: rewriting~\cite{RienerHMMS19,RienerLMM22},
  resubstitution~\cite{RienerTASM18,lee2021}, functional reduction, balancing, etc.
  \item \emph{Network transformation and mapping}: technology
  mapping~\cite{tempia2022}, graph mapping (network conversion)~\cite{tempia2022},
  LUT mapping, cleaning up networks, etc.
  \item \emph{Validation and verification}: CNF generation, miter
  generation, combinational equivalence checking, etc.
\end{itemize}

Many of the alogrithms in \emph{mockturtle} support taking user-defined 
custom cost functions. For example, for quantum and cryptography 
applications, the \emph{multiplicative complexity} can be used as
the optimization target, which considers only the cost of AND gates whereas
XOR gates are cost-free.

\subsection*{Example}
The following example code presents how \emph{mockturtle}'s various
algorithms can be combined to form a customized logic synthesis flow.

\begin{mdframed}
\begin{cppcode}
#include <mockturtle/mockturtle.hpp>
using namespace mockturtle;

/* Read in or create an AIG network */
aig_network aig = ...;

/* Map AIG into MIG using an exact database */
mig_npn_resynthesis resyn;
exact_library<mig_network> db( resyn );
mig_network mig = map( aig, db );

/* Optimize and clean up the network */
functional_reduction( mig );
mig = cleanup_dangling( mig );
mig_resubstitution( mig );
mig = cleanup_dangling( mig );

/* Create a miter network and check equivalence */
aig_network miter_ntk = *miter<aig_network>( aig, mig );
bool equivalent = *equivalence_checking( miter_ntk );

/* Write out the optimized MIG */
write_verilog( mig, "optimized.v" );
\end{cppcode}
\end{mdframed}

In line $5$, an initial benchmark circuit is created as an AIG. It can
be read in from a file using the parsing library \emph{lorina} (described
in more detail in Section~\ref{sec:lorina}) or be created manually. Then,
graph mapping is applied to convert the AIG into an MIG using an exact
NPN database (lines $8$-$10$). To optimize the MIG, we choose to apply
functional reduction (line $13$), which is applicable on all kinds of
networks, and a special version of resubstitution tailored for MIGs (line $15$).
We also check if the converted and optimized MIG is functionally equivalent 
to the original AIG (lines $19$-$20$). Finally, the resulting network
is be written out in Verilog format (line $23$).

\section{lorina: A parsing library}\label{sec:lorina}

The C++ library \emph{lorina} offers parsers for simple formats
commonly used in logic synthesis.  A parser reads a logic network in a
certain format from a file (or input stream) and invokes a callback
method of a visitor whenever the parsing of a primitive of the
respective format (e.g., an input, an output, or a gate definition) has been
completed.  These callback methods allow users to customize the
behavior of the parser and execute their code interleaved with the
parsing.  On parse error, a similar callback mechanism---the
\emph{diagnostic visitor}---is used to emit customizable diagnostics.

Each parser is implemented in its own header and provides a
\emph{reader function} \mintinline{cpp}{read_<format>} and a
\emph{reader visitor} \mintinline{cpp}{<format>_reader}, where
\mintinline{cpp}{<format>} has to be substituted by the name of the
respective format, e.g., \mintinline{cpp}{aiger},
\mintinline{cpp}{bench}, \mintinline{cpp}{blif},
\mintinline{cpp}{pla}, or \mintinline{cpp}{verilog}\footnote{A limited
  structural subset of Verilog is supported}.

The following example shows how to parse a two-level logic network
described as a programmable logic array from a file.

\begin{mdframed}
\begin{cppcode}
#include <lorina/pla.hpp>
using namespace lorina;

...

const auto r = read_pla("func.pla", pla_reader());
if (r == return_code::success)
{
  std::cout << "parsing successful" << std::endl;
}
else
{
  std::cout << "parsing failed" << std::endl;
}
\end{cppcode}
\end{mdframed}

A user can modify the default behavior of any parser by deriving a new
class from a reader visitor and overloading its virtual callback
methods.  Each method corresponds to an \emph{event point} defined by
the implementation of the parsing algorithm, e.g., the completion of
the parsing of the format's header information, or a certain input or
gate definition.

The listing below shows how to customize the \mintinline{cpp}{on_term}
event point of the reader visitor \mintinline{cpp}{pla_reader} such
that after a term is parsed, it is printed.  Note that the signatures
of the methods in derived classes have to exactly match their
counterparts in the base class.  The C++ keyword
\mintinline{cpp}{override} causes modern C++ compilers to warn on
signature mismatch and permits users to spot these errors quickly.

\begin{mdframed}
\begin{cppcode}
class reader : public pla_reader
{
public:
  void on_term(const std::string& term,
               const std::string& out) const override
  {
    std::cout << term << ' ' << out << std::endl;
  }
}; /* reader */
\end{cppcode}
\end{mdframed}

As a third parameter each reader function can optionally take a
diagnostic engine.  The engine is used to emit diagnostics when the
parsing algorithm encounters mistakes.  The possible error messages
are specified by the implementation of the parsing algorithm.

\begin{mdframed}
\begin{cppcode}
#include <lorina/diagnostics.hpp>

...

diagnostic_engine diag;
read_pla("func.pla", reader(), &diag);
\end{cppcode}
\end{mdframed}

Possible diagnostics for the PLA format could look as follows.

\begin{footnotesize}
\begin{verbatim}
[e] Unable to parse line
line 1: `i 16`
[e] Unsupported keyword `abc`
in line 4: `.abc`
\end{verbatim}
\end{footnotesize}

The diagnostic engine supports different levels of diagnostic
information and can emit one or multiple diagnostics depending on the
severity of the problem.  A diagnostic typically consists of a short
description of the problem and the line information to ease debugging.

Diagnostics can also be customized by overloading the
\mintinline{cpp}{emit} method as shown below.

\begin{mdframed}
\begin{cppcode}
class diagnostics : public diagnostic_engine
{
public:
  void emit(diagnostic_level level,
            const std::string& message) const override
  {
    std::cerr << message << std::endl;
  }
}; /* diagnostics */
\end{cppcode}
\end{mdframed}

\section{kitty: A truth table library}
The C++ library \emph{kitty} provides data structures and algorithms
for explicit truth table manipulation.  Truth table data structures and
algorithms are helpful for Boolean function manipulation, if the
functions are small, i.e., if they consist of up to 16 variables.
(For some algorithms, also functions with more variables can still be
efficiently manipulated.)  In such cases explicit truth table
representations can be significantly faster compared to symbolic
representations such as binary decision diagrams, since truth tables
require less overhead to manage than complicated data structures.  The
following listing is an example of how \emph{kitty} is used to create
truth tables that describe the two output functions of a full adder,
and to print them in hexadecimal format to the output.

\begin{mdframed}
\begin{cppcode}
#include <kitty/kitty.hpp>
using namespace kitty;

...

dynamic_truth_table a(3), b(3), c(3);

create_nth_var(a, 0);
create_nth_var(b, 1);
create_nth_var(c, 2);

const auto sum = a ^ b ^ c;
const auto carry = ternary_majority(a, b, c);

std::cout << "sum =   " << to_hex(sum) << "\n"
          << "carry = " << to_hex(carry) << "\n";
\end{cppcode}
\end{mdframed}

Inside the data structures, a truth table is represented in terms of
64-bit unsigned integers, called \emph{words}.  Each bit in a word
represents a function value.  For example, the truth table for the
function $x_0\land x_1$ is \texttt{0x8} (which is 1000 in base 2) and
the truth table for the majority-of-three function $\langle
x_0x_1x_2\rangle$ is \texttt{0xe8} (which is 11101000 in base 2).  A
single word can represent functions with up to 6 variables, since $2^6
= 64$.  A truth table for functions with 7 variables requires two
words, functions with 8 variables require four words, and so on.  In
general, an $n$-variable Boolean function, with $n \ge 6$, can be
represented using $2^{n-6}$ words.  On such truth table
representations, many operations for function manipulation can be
implemented using bitwise operations which map to efficient machine
instructions on a processor.  For a broader overview on how to
implement truth table operations using bitwise operations, we refer
the reader to the literature~\cite{Warren02,Knuth4A}.

The two main data structures for truth table manipulation in
\emph{kitty} are a static and a dynamic truth table.  The choice on
which to use depends on whether one knows the number of variables for
the function to represent at compile-time.  A static truth
table is more efficient at runtime, because it does not need to store
its number of variables and for many operations, the number of
iterations in a loop are compile-time constants.  For both data
structures, the number of variables is initialized when constructing
an instance and cannot be changed afterwards.  This avoids
reallocation of memory.  If the size of a truth table needs to be
changed, a new truth table must be created.  The previous example to
create the full adder functions uses dynamic truth tables.  Changing
the data type in Line~6 allows one to use a static instead of a
dynamic truth table; no other line must be changed:

\begin{mdframed}
\begin{cppcode*}{firstnumber=6}
static_truth_table<3> a, b, c;
\end{cppcode*}
\end{mdframed}

In addition to complete truth tables, which represent $n$-input
functions with $2^n$ bits, \emph{kitty} also has a \emph{partial 
truth table} data structure which holds arbitrary number of bits.
A partial truth table can be seen as a subset of the truth table of
a larger function that would be too inefficient to represent completely.
It can be used as a convenient container for bit streams supporting
bit parallel operations. Note that in partial truth tables, bit 
positions lose their correspondence to specific minterms.

\subsection*{Example}
We present a more complex example in which we first construct truth
tables from a Boolean chain (also called straight-line program or
combinational Boolean logic network) and then derive their algebraic
normal forms (also called positive-polarity Reed-Muller expression).
As input we use the implementation of an inversion in $\F_{2^4}$
described in~\cite[Fig.~1]{BP12}.  It can be represented as four
Boolean functions $y_i(x_1, x_2, x_3, x_4)$ for $1 \le i \le 4$.

\begin{mdframed}
\begin{cppcode}
std::vector<std::string> chain{
  "x5 = x3 ^ x4", "x6 = x1 & x3", "x7 = x2 ^ x6",
  "x8 = x1 ^ x2", "x9 = x4 ^ x6", "x10 = x8 & x9",
  "x11 = x5 & x7", "x12 = x1 & x4", "x13 = x8 & x12",
  "x14 = x8 ^ x13", "x15 = x2 & x3", "x16 = x5 & x15",
  "x17 = x5 ^ x16", "x18 = x6 ^ x17", "x19 = x4 ^ x11",
  "x20 = x6 ^ x14", "x21 = x2 ^ x10"};
\end{cppcode}
\end{mdframed}

Starting from the 4 primary inputs $x_1$, $x_2$, $x_3$, and $x_4$,
this chain assigns values to successive steps $x_5 = x_3 \oplus x_4$,
$x_6 = x_1 \land x_3$, $x_7 = x_2 \oplus x_6$, and so on.  Finally,
the functions representing $y_1$ to $y_4$ are computed by steps
$x_{18}$ to $x_{21}$.

\begin{mdframed}
\begin{cppcode*}{firstnumber=8}
std::vector<static_truth_table<4>> steps;

create_multiple_from_chain(4, steps, chain);

std::vector<static_truth_table<4>> y{
  steps[17], steps[18], steps[19], steps[20]};
\end{cppcode*}
\end{mdframed}

Note that the step indices in the \mintinline{cpp}{steps} vector are
off by 1, since indices start from 0.  Finally, we can print all truth
tables in hexadecimal representation, and also compute their algebraic
normal form and print the product terms they contain.

\begin{mdframed}
\begin{cppcode*}{firstnumber=14}
for (auto i = 0; i < 4; ++i)
{
  std::cout << "y" << (i + 1) << " = "
            << to_hex(y[i]) << "\n";

  const auto cubes = esop_from_pprm( y[i] );
  print_cubes(cubes, 4);
}
\end{cppcode*}
\end{mdframed}

The first lines of the output of this example program are as follows.
\begin{footnotesize}
\begin{verbatim}
y1 = af90
1-1-
-11-
-111
--1-
---1
...
\end{verbatim}
\end{footnotesize}
From the output one can readily obtain the algebraic normal form
$y_1 = x_1x_3 \oplus x_2x_3 \oplus x_2x_3x_4 \oplus x_3 \oplus x_4$.

\section{bill: A reasoning library}
The \emph{bill} library serves as an integration layer for symbolic
reasoning engines.  It supports conceptually different reasoning
approaches such as decision diagrams or satisfiability solvers and
provides simple and unified interfaces for integrating them.  The
design of the library is inspired by the architecture of metaSMT
framework~\cite{RHF+17} and focuses on header-only propositional
reasoning.

The following listing shows how a solving engine can be used to prove
De Morgan's law for propositional logic:
\begin{mdframed}
\begin{cppcode*}{firstnumber=1}
#include <bill/bill.hpp>

/* instantiate solving engine */
solver<solvers::ghack> solver;

/* construct problem instance */
auto const a = lit_type(solver.add_variable(),
                        lit_type::polarities::positive);
auto const b = lit_type(solver.add_variable(),
                        lit_type::polarities::positive);

auto const t0 =  add_tseytin_and(solver, a, b);
auto const t1 = ~add_tseytin_or(solver, ~a, ~b);
auto const t2 =  add_tseytin_xor(solver, t0, t1);
solver.add_clause(t2);

/* check for counterexamples using SAT */
auto const result = solver.solve()
if (result == result::states::unsatisfiable)
{
  std::cout << "proved" << std::endl;
}
else
{
  std::cout << "refuted" << std::endl;
}
\end{cppcode*}
\end{mdframed}

The listing consists of three parts:  First, a solving engine is
instances.  Second, the problem instance, i.e., De Morgan's law in
negated form is constructed in conjunctive normal form (CNF).  Third,
the solving engine is used to check if a counterexample to De Morgan's
law exists.  The unsatisifiability of the propositional formula
\begin{align*}
  (a \wedge b) \neq \neg(\neg a \vee \neg b)
\end{align*}
indicates that De Morgan's law is true for all valuations of $a$ and $b$.

Currently supported are ABC's \mintinline{cpp}{bsat} and
\mintinline{cpp}{bmcg} solvers and the MiniSAT variations Glucose,
MapleSAT, and GHack~\cite{ES03,AS15}.  Adding a new SAT solver to
\emph{bill} is as simple as declaring handful of interface functions.

\section{percy: An exact synthesis library}
The \emph{percy} library provides a collection of SAT-based exact synthesis
engines. These include engines based on conventional methods, as well as
state-of-the-art engines which can take advantage of DAG topology information
\cite{Knuth4A,HSMM19}.  The constraints and algorithms of such synthesis
engines may be quite dissimilar. Moreover, it is not always obvious which
combination  will be superior in a specific domain. It is often desirable to
experiment with several methodologies and solving backends to find the right
fit. The aim of \emph{percy} is to provide a flexible common interface that
makes it easy to construct a parameterizable synthesis engine suitable for
different domains.

The \emph{percy} library also serves as an example of the ideas presented in
this paper.  It is built on top of \emph{kitty}, which it uses to construct
synthesis specifications.  Thus, it shows how the lightweight libraries proposed
here can be easily composed to build up ever more complex structures.

Synthesis using \emph{percy} concerns five main components:
\begin{enumerate}
\item \emph{Specifications} -- Specification objects contain the
  information essential to the synthesis process such as the functions to
  synthesize, I/O information, and a number of optional parameters such
  as conflict limits for time-bound synthesis, or topology information.
\item \emph{Encoders} -- Encoders are objects which convert specifications
  to CNF formul\ae.  There are various ways to create such encodings, and
  by separating their implementations it becomes simple to use
  encodings in different settings.
\item \emph{Solvers} -- Once an encoding has been created, we use a
  SAT solver to find a solution.  Various different SAT solving backends are supported.
\item \emph{Synthesizers} -- Synthesizers perform the task of composing
  encoders and solvers.  Different synthesizers correspond to different
  synthesis flows.  For example, some synthesizers may support synthesis
  flows that use topological constraints, or allow for parallel synthesis
  flows.  To perform synthesis using \emph{percy}, one creates a
  synthesizer object.  This object can then be parameterized by changing
  settings such as its encoder or solver backends.
\item \emph{Chains} -- Boolean chains are the result of exact synthesis.  A
  Boolean chain is a compact multi-level logic representation that can
  be used to represent multi-output Boolean functions.
\end{enumerate}
A typical workflow will have some source for generating
specifications, which are then given to a synthesizer that converts
the specifications into optimum Boolean chains. Internally, the
synthesizer will compose its underlying encoder and SAT solver in its
specific synthesis flow.  For example in \cite{RHM+19}, a resynthesis
algorithm generates cuts in a logic network which serve as
specifications.  They are then fed to a synthesizer, and if the
resulting optimum Boolean chains lead to an improvement, are replaced
in the logic network.  In optimizing this workflow, \emph{percy} makes
it easy to swap out one synthesis flow for another, to change CNF
encodings, or to switch to a different SAT solver.

\subsection*{Example}
In the following example, we show how \emph{percy} can be used to synthesize
an optimum full adder.  While simple, the example shows some common interactions
between the components.

\begin{mdframed}
\begin{cppcode}
#include<percy/percy.hpp>
using namespace percy;
using namespace kitty;

...

/* start by creating the functions to synthesize */
static_truth_table<3> x, y, z;

create_nth_var( x, 0 );
create_nth_var( y, 1 );
create_nth_var( z, 2 );

auto const sum = x ^ y ^ z;
auto const carry = ternary_majority(x, y, z);

/* create a specification using the functions */
synth_spec<static_truth_table<3>> spec;
spec.nr_in = 3;
spec.nr_out = 2;
spec.functions[0] = sum;
spec.functions[1] = carry;

/* instantiate a synthesizer and find an optimum chain */
chain c;
auto const res = synthesize( spec, c );

/* verify that the chain is functionally correct */
if ( res == percy::success )
{
  auto const out_functions = c.simulate;
  assert( out_functions[0] == sum );
  assert( out_functions[1] == carry );
}
\end{cppcode}
\end{mdframed}

The \mintinline{cpp}{synthesize} function takes five parameters.  In
the example above, only two of them are set the specification
(\mintinline{cpp}{spec}) and the chain to store the synthesis result
(\mintinline{cpp}{c}).  The other three parameters are optional and
can be used to select a specific backend for SAT solving, an encoder,
and the synthesis method.  By default, the standard synthesis engine
(\mintinline{cpp}{SYNTH_STD}) is invoked together with ABC's
\mintinline{cpp}{bsat} solver.  Further, by default the specification
is encoded using the SSV encoding (see \cite{HSMM19} for details about
the different supported encodings).  Suppose that that this particular
combination is not suitable for a workflow, then we can easily switch
to a different encoding or solving backend by changing only one line
of code:

\begin{mdframed}
\begin{cppcode*}{firstnumber=26}
  auto const res = synthesize( spec, c, SLV_GLUCOSE,
                               ENC_DITT, SYNTH_STD_CEGAR );
\end{cppcode*}
\end{mdframed}

In doing so we switch to a synthesis engine which synthesizes Boolean
chains, with the DITT encoding, using a synthesis method based on
counterexample-guided abstraction-refinement (CEGAR), and Glucose SAT
solver as backend.  While we now use a completely different synthesis
engine, its interface remains the same.

\section{easy: An ESOP library}

The C++ library \emph{easy} provides implementations of verification
and synthesis algorithms for \emph{exclusive-or sum-of-product} (ESOP)
forms. An ESOP form is a two-level logic representation that consists
of one level of multi-fanin AND-gates, followed by one level of
multi-fanin XOR-gates. For instance, the Boolean expression
\[
  x_1\bar x_2\bar x_3\bar x_4 \oplus
  x_0\bar x_3x_4 \oplus
  x_0x_2\bar x_3 \oplus
  x_0x_1x_2x_3 \oplus
  1
\]
is an ESOP form that realizes the Boolean function
\texttt{0xcafeaffe}.

ESOP forms have good testability properties and allow for a compact
representation of arithmetic circuits. The inherent reversibility of
the XOR-gate, moreover, sparked interest in ESOP forms in application
fields like cryptography and quantum computation.

The \emph{easy} library implements truth table based algorithms using
\emph{kitty} to verify that an ESOP form realizes a completely- or
incompletely-specified Boolean function or to verify that two ESOP
forms are functionally equivalent.  Being truth table based, these
algorithms are particularly effective when Boolean functions with $16$
or less Boolean variables are represented.

The ESOP representation is not canonical.  One Boolean function can be
expressed by multiple structurally different, but semantically
equivalent ESOP forms.  For many applications, ESOP forms with small
(or minimum) costs with respect to a cost criterion, e.g., the number
of product terms or the number of AND-gates, are of interest. The
\emph{easy} library provides various heuristic and exact methods for
synthesizing ESOP forms from a given Boolean function:

\begin{enumerate}
\item \emph{Decomposition based ESOP synthesis methods} recursively
  decompose the Boolean function and re-compose its ESOP form from the
  individual parts.  These methods are fast, but in general do not lead
  to an ESOP form of minimum size.  Within easy, decomposition
  algorithms using \emph{Pseudo-Kronecker Reed-Muller} (PKRMs) and
  \emph{Positive Polarity Read-Muller} (PPRMs) forms are implemented,
  which are both special cases of ESOP forms.
\item \emph{SAT based ESOP synthesis methods} formulate the problem of
  synthesizing an ESOP form with a fixed number of $k$ product terms
  as a constraint satisfaction problems.  A satisfying assignment for
  the constraint system directly corresponds to a realization of an
  ESOP from with $k$ product terms. The \emph{easy} library supports
  two formulations of SAT based ESOP synthesis---one based on the
  Helliwell equation~\cite{MSE+19}, the other based on Boolean
  learning~\cite{RES+18}. The latter formulation uses a
  counterexamples-guided abstraction-refinement loop to cope with many
  Boolean variables and large don’t care sets. In an iterative search
  procedure, these synthesis algorithms compute ESOP forms of
  minimum size.
\end{enumerate}

\section{caterpillar: A quantum circuit synthesis library}
\emph{Caterpillar} is a library dedicated to the problem of synthesizing circuits for fault-tolerant quantum computing. 
This application usually deals with many error-corrected qubits and the cost of circuit is usually measured in terms of number of $T$ gates. 

The library focuses on the synthesis of quantum circuits starting from logic network representations of the desired function and it aims at minimizing the number of gates and qubits generated.
One relevant application of the library is the synthesis of oracle circuits for large Boolean functions. 
In addition to the synthesis of classical functions, the library can be also used to compose pre-optimized quantum operations, with the guarantee of fitting the targeted hardware. 

\emph{Caterpillar} interfaces with all the logic networks defined in the library \emph{mockturtle}, e.g., AIG, XAG, LUT, etc., and uses \emph{tweedledum} to describe quantum circuits and to decompose reversible gates into quantum gates.

\subsection{Logic Network Synthesis}
The core algorithm of the library is a hierarchical method called \textit{logic network synthesis}.
The algorithm takes any classical logic network as input and returns a reversible network, saving all the intermediate results on helper qubits, also called \textit{ancillae}. 
The library also provides an LUT mapping method designed specifically for this application. This mapping allows us to decompose the initial network into LUT, enabling control over the size of each reversible gate and the number of generated \textit{ancillae}. Further details can be found in~\cite{MSR+19}.

\subsection{Mapping strategies}
The synthesis task includes uncomputing all intermediate results. The easiest strategy to do so is the \emph{Bennett strategy} in which the obtained reversible circuit uses one \textit{ancilla}, for each node in the network. Nevertheless, \emph{caterpillar} supports several different uncomputing methods, called \emph{strategies}. 

\subsection*{Bennett strategy}
A strategy that consists in computing all the nodes in topological order and uncomputing them in inverse topological order. It has been described in~\cite{Bennett89} and it provides a solution that always returns the smallest number of reversible gates and the highest number of ancillae, with respect to the other methods.
\subsection*{Eager strategy}
This strategy computes each node in topological order. Once it reaches a primary output, all nodes in the transitive fanin are uncomputed as they are no longer required.
This strategy does not support in-place operations.
\subsection*{Best-fit strategy}
This strategy applies a $k$-LUT mapping that decomposes the network into cells. Cells are initially placed in the reversible network following an eager strategy. The logic contained into each cell is decomposed by a second $k$-LUT mapping. The method selects the minimum $k$ such that there are enough clean ancillae to save intermediate results. A different “best-fit” $k$ is selected for each cell. Further details can be found in~\cite{MSR+18}.
\subsection*{Pebbling strategy}
The pebbling strategy is obtained by solving iteratively the reversible pebbling game on the given network. The problem is encoded as a SAT problem and addressed by state-of-the-art solvers. This strategy allows us to explore the trade-off between qubits and operations. Details can be found in~\cite{MSRB+19}.
\subsection*{XAG strategy}
This strategy is a constructive method dedicated to XAG graphs. It is based on two observations: the XOR operation is relatively inexpensive in fault tolerant quantum computing and 
Toffoli gates used to implement AND nodes can be uncomputed using 0 $T$ gates. The resulting circuit achieves a number of expensive gate that is proportional to the number of AND nodes in the initial network.
Details can be found in~\cite{MSC+19}.

\subsection*{Example}
In this example we read an XAG network from a verilog file and perform logic network based synthesis to obtain the corresponding reversible circuit. 
We show how to use the pebbling mapping strategy to constraint the number of available qubits.
\begin{mdframed}
\begin{cppcode}
#include <caterpillar/caterpillar.hpp>
#include <lorina/verilog.hpp>
#include <mockturtle/io/verilog_reader.hpp>
#include <mockturtle/networks/xag.hpp>
#include <tweedledum/networks/netlist.hpp>

using namespace caterpillar;
using namespace mockturtle;
using namespace lorina;
using namespace tweedledum;

...

/* read verilog file using lorina */
xag_network xag;

auto const res = read_verilog("ex.v", verilog_reader(xag));
if (res != return_code::success)
	return;

/* set parameters of the pebbling compilation strategy*/
pebbling_mapping_strategy_params ps;
ps.pebble_limit = 100;
pebbling_mapping_strategy<xag_network> strategy(ps);

/* synthesize the quantum circuit */
netlist<caterpillar::stg_gate> circ;
logic_network_synthesis(circ, xag, strategy);
\end{cppcode}
\end{mdframed}

\section{Angel: A Quantum State Preparation Library}
\emph{angel} is a modern library for \emph{Quantm State Preparation}~(QSP). The \emph{angel} library implements algorithms with the purpose of synthesizing an optimized quantum circuit to prepare a given quantum state.  As an objective function, the algorithms focus on minimizing the number of control qubits and thereby the circuit's depth and the number of elementary quantum gates.  In particular, the algorithms reduce the number of controlled-NOT gates, which are in many experimental NISQ architectures relatively expensive when compared to other elementary quantum gates.  

Finding the optimum quantum circuit with the minimum number of elementary gates, however, is in practice for arbitrary quantum states intractable.  Hence,  we tried to identify families of quantum states that can be prepared efficiently and precisely.  In the current version, \emph{angel} implements algorithms for preparing uniform quantum states,  and sparse quantum states.

\subsection{Uniform quantum state preparation}
These states are superpositions of basis states, where all amplitudes are either zero or have the same value.  As a key point, we map uniform quantum states to Boolean functions.  Representing uniform quantum states as Boolean functions allows us to employ the Shannon decomposition to solve the state preparation problem recursively.  Our algorithm iterates over the variables of the Boolean function, which correspond to qubits, and prepares them one by one, by computing the probability of being zero for the variable depending on previously prepared variables.  This computational step requires counting the number of ones for each recursive co-factor of the Boolean function.  The probability is then the number of ones of the current function divided by the number of ones of the negative co-factor.  To reduce the number of elementary quantum gates, we utilize decision diagrams and functional dependency analysis.

\textbf{Using Decision Diagrams.}
We have presented an implementation of this algorithm in~\cite{Mozafari20} using \emph{Binary Decision Diagrams}~(BDDs) as a representation of Boolean functions and dynamic programming.  BDDs are particularly suitable for our purpose because counting and co-factoring can be very efficiently implemented as BDD operations. More details are available in~\cite{Mozafari20}.

\textbf{Using Functional Dependencies.}
Utilizing Boolean functions allows us to identify functional dependencies among variables.  We make use of variable reordering and dependency analysis methods. Both methods can be either implemented as exact algorithms or heuristics, which allows us to choose between different runtime-quality trade-offs. The exact algorithms generally take more runtime with better results while heuristics close to the answer with fewer runtime. The detailed structure is presented in~\cite{mozafari2021efficient}.

\subsection{Sparse quantum state preparation}
These states are superpositions over a given set $S \subset \{0,1\}^n , |S|<<2^n$.  We present these states using \emph{Algebraic Decision Diagram} (ADD).  Each path in the ADD shows its corresponding basis state from the set $S$. We go through each path and prepare each basis state efficiently.  Then, we use an ancilla qubit to show the basis state is prepared.  As an alternative, we use BDDs to prepare sparse and uniform quantum states.

\subsection{Examples}
We show how to use a dependency analysis and variable reordering algorithm to synthesize a quantum circuit from a Boolean function given as a truth table using \emph{angel} in combination with \emph{kitty} and \emph{tweedledum}.

\begin{mdframed}
\begin{cppcode}
#include <angel/angel.hpp>
#include <tweedledum/IR/Circuit.h>
#include <kitty/kitty.hpp>

/* Prepare a truth table */
kitty::dynamic_truth_table tt( 3 );
kitty::create_from_binary_string( tt,
  std::string( "1000" "0001" ) );
  
/* Prepare tweedledum's network type */
tweedledum::Circuit network;

/* Setup ESOP-based dependency analysis */
angel::esop_deps_analysis::parameter_type epars;
angel::esop_deps_analysis::statistics_type estats;
angel::esop_deps_analysis esop( epars, estats );

/* Setup exhaustive reordering strategy */
angel::exhaustive_reordering order;

/* Prepare parameters and statistics */
angel::state_preparation_parameters ps;
angel::state_preparation_statistics st;

/* Perform state preparation */
angel::uniform_qsp_deps
  <decltype(network), decltype( esop ), decltype( order )>
  ( network, esop, order, tt, ps, st);
  
\end{cppcode}
\end{mdframed}

In the lines $6$ to $8$, we prepare the truth table of a $3$-input Boolean function using \emph{kitty}.
In the lines $14$ to $16$, we setup the dependency strategy, which takes {\tt epars} and {\tt estats} as arguments.  The parameters are inputs and can be used to customize the dependency analysis algorithms.  The statistics are outputs generated by the algorithm containing runtime as well as the number and types of identified dependencies.
In line $19$, we setup the reordering strategy, which considers all possible reordering of the $3$-input Boolean function.  Other strategies, such as random or greedy reordering, exist.
In line $26$, we call our algorithm with the setup dependency and reordering strategies to prepare the quantum state given as a truth table. The final circuit is available as a network from \emph{tweedledum}.

We show how to synthesize a quantum circuit for an sparse quantum state  using \emph{angel} in combination with \emph{CUDD}~\cite{somenzi2009cudd} and \emph{tweedledum}.

\begin{mdframed}
\begin{cppcode}
#include <angel/angel.hpp>
#include <tweedledum/IR/Circuit.h>
#include <cudd/cudd.h>
#include <cudd/cuddInt.h>

/* Create ADD from a map of 
	            basis states and their amplitudes */
Cudd cudd;
auto const f_add = create_add( cudd, map_amplitudes );

/* Prepare tweedledum's network type */
tweedledum::Circuit network;

/* Prepare statistics */
angel::sparse_qsp_statistics stats;

/* Perform state preparation  */
angel::sparse_qsp<network_type>( network, f_add, stats );
\end{cppcode}
\end{mdframed}

In the lines $8$ to $9$, we read a map consisting of basis states and their corresponding amplitudes and use CUDD to create its corresponding ADD. In the lines $12$, we specify returning network type from \emph{tweedledum}. In line $15$, we setup the variable {\tt stats} that results in the final statistics of our generated circuit. Finally, in line $18$, we call our QSP algorithm.

\section{Example: Mining optimum logic networks}
The paper concludes with the example \emph{exactmine} from the
showcase repository.  The example implements a program---with the help
of four presented libraries---that can mine optimum networks for
truth tables.  It uses \emph{kitty} to manage truth tables for which
optimum networks are found using \emph{percy}.  Truth tables can be
entered manually or extracted from LUT (lookup-table) networks using
\emph{lorina}.  Finally, all functionality is exposed to the user in
terms of an \emph{alice} shell.  The complete source code is available
in the repository.  Less important aspects are omitted.  To save
space, we also omitted all namespace prefixes for the logic synthesis
libraries.

A typical \emph{exactmine} session is as follows:

\begin{footnotesize}
\begin{verbatim}
exactmine> help
Exact synthesis commands:
 find_network

Loading commands:
 load        load_bench

General commands:
 alias       convert     current      help
 print       ps          quit         set
 show        store
exactmine> load cafe
exactmine> load affe
exactmine> set npn 1
exactmine> load_bench -t 3 adder.bench
exactmine> store -o
[i] networks in store:
     0: cafe
     1: affe
     2: 6
     3: 1e
     4: 01
     5: 69
     6: 07
     7: 1
     8: 06
  *  9: 17
exactmine> current -o 0
exactmine> find_network
exactmine> current -o 1
exactmine> find_network --verify
[i] synthesized chain matches specification
exactmine> store -o
[i] networks in store:
     0: cafe, optimum network computed
  *  1: affe, optimum network computed
     2: 6
     3: 1e
     4: 01
     5: 69
     6: 07
     7: 1
     8: 06
     9: 17
exactmine> print -o
function (hex): affe
function (bin): 1010111111111110
optimum network: {a{(b!d)[cd]}}
\end{verbatim}
\end{footnotesize}

Command \ilcmd{help} lists all commands in the shell.  Besides the
default \emph{alice} commands, three custom commands are implemented
in \emph{exactmine}: \ilcmd{load} to load a truth table into the
store, \ilcmd{load\_bench} to load LUTs from a BENCH file, and
\ilcmd{find\_network} to find an optimum network for a store element.
First two truth tables are entered explicitly, and afterwards all LUT
functions that do not exceed 3 inputs are extracted from a LUT network
in BENCH format.  Setting the shell variable \emph{npn} to 1 will
insert the NPN class of a function into the store instead of the
function itself.  No duplicate is added to the store.  The first store
element is selected using the \ilcmd{current} command, before an
optimum network is computed for it.  The same is repeated for the
second store element.  Another output of the store elements using \ilcmd{store -o}
confirms that now the first two store elements have an associated
optimum network, which is printed for the current network in the last
command.

\subsection*{Store type}
The \emph{alice} shell contains a single store type for optimum
networks, which is a pair of a truth table and an expression.

\begin{mdframed}
\begin{cppcode}
class optimum_network
{
public:
  /* constructors */
  bool exists() const
  {
    const auto num_vars = function.num_vars();

    /* a global hash table for each number of variables */
    static std::vector<std::unordered_set<
      dynamic_truth_table,
      hash<dynamic_truth_table>>> hash;

    /* resize hash tables? */
    if (num_vars >= hash.size())
    {
      hash.resize(num_vars + 1);
    }

    /* insert into hash table */
    const auto r = hash[num_vars].insert(function);

    /* did it exist already? */
    return !r.second;
  }

public: /* field access */
  dynamic_truth_table function{0};
  std::string network;
};
\end{cppcode}
\end{mdframed}

The store type has a method \mintinline{cpp}{exists} that accesses a
global hash table to check whether the function has already been
computed, or inserts it into the truth table, if it does not exist
already.  This function is being used by the \ilcmd{load...} commands
to avoid duplicates in the store.

The following code registers the store type using the access flag
\ilcmd{-o} to alice and implements the functionality for \ilcmd{store
  -o} and \ilcmd{print -o}:

\begin{mdframed}
\begin{cppcode*}{firstnumber=33}
ALICE_ADD_STORE(optimum_network, "opt", "o", ...)

ALICE_DESCRIBE_STORE(optimum_network, opt)
{
  if (opt.network.empty())
  {
    return to_hex(opt.function);
  }
  else
  {
    return format("{}, optimum network computed",
                  to_hex(opt.function));
  }
}
\end{cppcode*}
\end{mdframed}
\begin{mdframed}
\begin{cppcode*}{firstnumber=47}
ALICE_PRINT_STORE(optimum_network, os, opt)
{
  os << format("function (hex): {}\nfunction (bin): {}\n",
               to_hex(opt.function),
               to_binary(opt.function));

  if (opt.network.empty())
  {
    os << "no optimum network computed\n";
  }
  else
  {
    os << format("optimum network: {}\n", opt.network);
  }
}
\end{cppcode*}
\end{mdframed}

\subsection*{Load truth tables}
We now describe the two commands \ilcmd{load} and \ilcmd{load\_bench}
to load truth tables into the store.  Both use a common function,
which computes the NPN class representations of the truth table, if the alice variable
\emph{npn} is set and also checks whether the truth table is already
in the store.

\begin{mdframed}
\begin{cppcode*}{firstnumber=62}
void add_optimum_network_entry(command& cmd,
        dynamic_truth_table& func)
{
  /* compute NPN? */
  if (cmd.env->variable("npn") != "")
  {
    func = std::get<0>(exact_npn_canonization(func));
  }

  /* add to store if it does not exist yet */
  optimum_network entry(func);
  if (!entry.exists())
  {
    cmd.store<optimum_network>().extend();
    cmd.store<optimum_network>().current() = entry;
  }
}
\end{cppcode*}
\end{mdframed}

The \ilcmd{load} commands loads a truth table from user input in
hex format.

\begin{mdframed}
\begin{cppcode*}{firstnumber=79}
class load_command : public command
{
public:
  load_command( const environment::ptr& env )
      : command( env, "Load new entry" )
  {
    add_option("truth_table,--tt", table,
               "truth table in hex format");
  }

  protected:
  void execute() override
  {
    unsigned num_vars = ::log(table.size()*4) / ::log(2.0);
    dynamic_truth_table func(num_vars);
    create_from_hex_string(func, truth_table);
    add_optimum_network_entry(*this, func);
  }

  private:
  std::string table;
};

ALICE_ADD_COMMAND(load, "Loading");
\end{cppcode*}
\end{mdframed}

The \ilcmd{load\_bench} command extracts all truth tables in a LUT
network in BENCH format.  The maximum size of the LUTs can be controlled using
a threshold parameter.

\begin{mdframed}
\begin{cppcode*}{firstnumber=103}
class load_bench_command : public command
{
public:
  /* constructor for argument parsing */
  ...

  class lut_parser : public bench_reader
  {
  public:
    lut_parser(load_bench_command& cmd) : cmd(cmd) {}

    void on_gate(const std::vector<std::string>& inputs,
                 const std::string& output,
                 const std::string& type) const override
    {
      const auto num_vars = inputs.size();

      if (num_vars > cmd.threshold) return;

      dynamic_truth_table func(num_vars);
      create_from_hex_string(func, type.substr(2u));
      add_optimum_network_entry(cmd, func);
    }

  private:
    load_bench_command& cmd;
  };

protected:
  void execute() override
  {
    read_bench(filename, lut_parser(*this));
  }

private:
  std::string filename;
  unsigned threshold = 6u;
};

ALICE_ADD_COMMAND(load_bench, "Loading");
\end{cppcode*}
\end{mdframed}

\subsection*{Find optimum networks}
The \ilcmd{find\_network} command takes the current store element and
compute an optimum network using exact synthesis. One can request to check
whether the synthesized network realizes the input function using the argument
\ilcmd{-{}-verify}.  The command implements validity rules, which are checked
before the command is executed.  The rules ensure that a current store element
is available, and that the current store element has not yet an optimum network
assigned to it (unless overridden by the \ilcmd{-f} argument).

\begin{mdframed}
\begin{cppcode*}{firstnumber=143}
class find_network_command : public command
{
public:
  find_network_command(const environment::ptr& env)
      : command(env, "Find optimum network")
  {
    add_flag("--verify", "..." );
    add_flag("--force,-f", "..." );
    add_flag("--verbose,-v", "..." );
  }
protected:
  rules validity_rules() const override
  {
    return {
      has_store_element<optimum_network>( env ),
      {[this]() {
        auto opt = store<optimum_network>().current();
        return opt.network.empty() || is_set("force");
      },
      "network already computed (use -f to override)"}
    };
  }
\end{cppcode*}
\end{mdframed}
\begin{mdframed}
\begin{cppcode*}{firstnumber=165}
  void execute() override
  {
    auto& opt = store<optimum_network>().current();

    synth_spec<dynamic_truth_table> spec;
    spec.nr_in = opt.function.num_vars();
    spec.nr_out = 1;
    spec.verbosity = is_set("verbose") ? 1 : 0;
    spec.functions[0] = &opt.function;

    auto synth = new_synth(spec, type);
    chain<dynamic_truth_table> c;

    if (synth->synthesize(spec, c) != success)
    {
      env->out() << "[e] could not find optimum network\n";
      return;
    }

    if (is_set("verify"))
    {
      if (*(c.simulate())[0] == opt.function)
      {
        env->out() << "[i] synthesized chain matches "
                      "specification\n";
      }
      else
      {
        env->err() << "[e] synthesized chain does "
                      "not match specification\n";
        return;
      }
    }
    std::stringstream str;
    c.to_expression( str );
    opt.network = str.str();
  }
};

ALICE_ADD_COMMAND(find_network, "Exact synthesis")
\end{cppcode*}
\end{mdframed}

\section{Acknowledgments}
We like to thank Alan Mishchenko for inspiring this project.  We also
thank Luca Amar\`u for helpful discussions and code contributions.
Finally, we thank Synopsys and all reviewers of previous versions of this paper.
This research was supported by the EPFL Open Science Fund and the Swiss
National Science Foundation (200021-169084 MAJesty, 200021-1920981 Supercool).

\bibliographystyle{IEEEtran}
\bibliography{library,ref}

\end{document}